\documentstyle[12pt,world_sci]{article}

\def\D0{D$\!$\O }

\newcommand{\rstev}{\mbox{$\sqrt{s}$ =\ 1.8\ TeV }}
\newcommand{\AP}{\mbox{${\rm\bar{p}}$ }}
\newcommand{\PAP}{\mbox{{\rm p}\AP}}
\newcommand{\ET}{\mbox{$E_{T}$ }}

\newcommand{\MET}{\mbox{$\not$E$_{T}$ }}
\newcommand{\PT}{\mbox{$P_{T}$ }}
\def\gw{\mbox{$\Gamma(W)$}}
\def\gz{\mbox{$\Gamma(Z)$}}
\def\gmw{\mbox{$\Gamma(W \rightarrow l\nu)$}}
\def\gmz{\mbox{$\Gamma(Z \rightarrow l^+l^-)$}}
\def\sgw{\mbox{$\sigma(\PAP \rightarrow W + X)$}}
\def\sgz{\mbox{$\sigma(\PAP \rightarrow Z + X)$}}

\newcommand{\lum}{\mbox{$ \int{Ldt} $}}
\newcommand{\ipb}{\mbox{$ {\rm pb}^{-1} $}}
\newcommand{\sbr}{\mbox{$ \sigma\cdot{\rm B} $}}
\newcommand{\nb}{\mbox{nb}}
\newcommand{\Wen}{\mbox{$ W \rightarrow e \nu_e$ }}
\newcommand{\Wmn}{\mbox{$ W \rightarrow \mu \nu_{\mu}$ }}
\newcommand{\Wte}{\mbox{$ W \rightarrow \tau \rightarrow e$ }}
\newcommand{\Wtm}{\mbox{$ W \rightarrow \tau \rightarrow \mu$ }}
\newcommand{\Wln}{\mbox{$ W \rightarrow l \nu$ }}
\newcommand{\Zee}{\mbox{$ Z \rightarrow {e^+e^-}$ }}
\newcommand{\Zmm}{\mbox{$ Z \rightarrow {\mu^+\mu^-}$ }}
\newcommand{\Ztt}{\mbox{$ Z \rightarrow {\tau^+\tau^-}$ }}

\newcommand{\Zll}{\mbox{$ Z \rightarrow {l^+l^-}$ }}
\begin{document}

\pagestyle{empty}


\title{MEASUREMENT OF $W$ AND $Z$ PRODUCTION
CROSS-SECTIONS IN \PAP COLLISIONS AT \rstev}
\author{PAUL Z. QUINTAS\thanks{Representing the \D0\ Collaboration.}\\
{\em Fermi National Accelerator Laboratory \\ Batavia, IL 60510}}
\maketitle

\begin{abstract}
The cross sections for $W$ and $Z$ production
in \PAP collisions at \rstev\\
are measured using the \D0 detector
at the Fermilab Tevatron collider.  The detected final states are
\Wen, \Zee, \Wmn, and \Zmm.
In the ratio of these measurements,
many common sources of systematic error cancel
and we measure
$R=\sigma (\PAP\ \rightarrow W) \cdot Br(W\rightarrow l\nu)$/
$\sigma (\PAP\ \rightarrow Z) \cdot Br(Z\rightarrow l^+l^-)$.
Assuming standard model couplings, this result is used to
determine the width of the $W$ boson
and to set a limit on the
decay $W^+\rightarrow t \bar b$.
\end{abstract}

\section{Introduction}

The cross-sections times branching ratios for $\PAP \rightarrow W$ and
$\PAP \rightarrow Z$ with decays into final states
with electrons or muons
are measured using the \D0 detector.
In the ratio of these measurements,
many common sources of systematic error cancel
and we measure
\[ R= \frac{\sgw \cdot B(W \rightarrow l\nu)}
           {\sgz \cdot B(Z \rightarrow l^+l^-)}. \]
This ratio is of interest since it can be expressed as
the product of calculable or well-measured quantities:
\[ R =
\frac{\sgw}{\sgz} \hskip3mm \frac{\gmw}{\gmz} \hskip3mm \frac{\gz}{\gw}. \]
This gives the most precise measurement of \gw\ .
A width measurement which exceeds the standard model value might indicate
non-standard decays of the $W$.

\section{The \D0\ Detector}

The \D0\ detector \cite{D0NIM} consists of three major subsystems:
central tracking detectors, nearly hermetic liquid argon calorimetry,
and a muon spectrometer. The central tracking system is used to
identify tracks in the psuedorapidity range $|\eta| \le 3.5$.
The calorimeter covers the region up to $|\eta| \le 4$
with energy resolution for electrons approximately $15\%/\sqrt{E}$.
The muon system consists of drift chambers and magnetized iron toroids.

\section{Electron Channel Cross-sections}

The \Wen and \Zee candidates were collected on a single trigger
consisting of a two levels: the hardware trigger required
electromagnetic energy above threshold (usually 10 GeV)
in a .2 x .2 ($\eta$ x $\phi$) tower.
The software trigger required  \ET $\ge$ 20 GeV
and made loose shower shape and isolation cuts.

Offline electrons were required to have mostly electromagnetic
energy, a shower shape consistent transversely and longitudinally
with an electron, be isolated from other activity in the calorimeter,
and be matched with a central detector track.
Fiducial cuts of         $|\eta| \le 1.1$
              or $1.5 \le |\eta| \le 2.5$
ensured a good trigger, good energy resolution, and low background.

The \Wen candidates satisfied $\ET >$ 25 GeV
and $\MET >$ 25 GeV, resulting in a sample of 10346 candidates.
The \Zee candidates had two electrons with $\ET \ge 25$ GeV,
one of which satisfied all the \Wen cuts, and the other
passed all except the track match cut. There was
an invariant mass cut of $75 \le M_{inv} \le 105$ GeV,
creating a sample of 782 candidates.

Electron efficiencies are determined from
the \Wen sample with a harder \MET cut and from the \Zee sample.
The \Wen backgrounds are estimated separately
either from data (QCD) or Monte Carlo (\Wte and \Zee).
The \Zee background is estimated by fitting the invariant mass peak
to a Breit-Wigner convoluted with the detector resolutions
plus a linear background.
Luminosity for this trigger was $\lum = 12.4 \pm 1.5 \; \ipb$.
Table \ref{tab:xsecs} gives the preliminary values of the cross-sections.

\begin{table}[hbt]
\caption{Measured $W$ and $Z$ Cross-sections}
\begin{tabular}{|c|c|c|c|c|}  \hline
	        & \Wen              & \Zee
	        & \Wmn              & \Zmm           \\ \hline
Number of Events  & 10346           & 782
		  & 1665           & 77         \\ \hline \hline
Acceptance	& $46.1 \pm 0.9$ \% & $36.4 \pm 0.5$ \%
		& $25.1 \pm 0.7$ \% & $6.7 \pm 0.4$  \% \\ \hline
Trig + Sel Eff. & $73.7 \pm 1.8$ \% & $74.6 \pm 3.0$ \%
		& $22.4 \pm 2.6$ \% & $53.8 \pm 5.0$ \% \\ \hline \hline
Total Bkgd      & $ 5.9 \pm 0.7$ \% & 5.2 $\pm$ 2.3 \%
		& $22.1 \pm 1.9$ \% & 10.1 $\pm$ 3.7 \% \\ \hline \hline
Luminosity      & \multicolumn{2}{c|}{$12.4 \pm 1.5 \ipb$}
		& \multicolumn{2}{c|}{$11.1 \pm 1.3 \ipb$} \\ \hline \hline
\sbr (\nb)        &       2.32     & 0.220
		  &       2.09     & 0.174       \\
Stat. Err.        & $\pm$ 0.02     & $\pm$ 0.008
		  & $\pm$ 0.07     & $\pm$ 0.022 \\
Syst. Err.        & $\pm$ 0.07     & $\pm$ 0.011
		  & $\pm$ 0.22     & $\pm$ 0.018 \\
Lum. Err.         & $\pm$ 0.28     & $\pm$ 0.026
		  & $\pm$ 0.25     & $\pm$ 0.021 \\ \hline
\end{tabular}
\label{tab:xsecs}
\end{table}

\section{Muon Channel Cross-sections}

The \Wmn and \Zmm trigger consisted of
two hits-in-road searches at the hardware level
(first a coarse road, then a fine road,
with an effective \PT cut of 7 GeV).
The software trigger did track finding and reconstruction
and required at least one muon with \PT $\ge$ 15 GeV.

Offline tracks (``loose'' muons) were confirmed by energy in the calorimeter
and required to pass through the central iron ($|\eta| \le 1.0$)
and through a minimum magnetic field ($\int B \cdot dl \ge 2 \; Tm$).
``Tight'' muons were defined satisfy the loose requirements plus
have a matching track in the central detector,
have a good quality fit to the vertex point,
central detector track and muon track,
be in time with beam crossing,
and be isolated in the calorimeter.

The \Wmn candidates have a tight muon and
satisfied kinematic cuts of $\PT \ge 20$ GeV and $\MET \ge 20$ GeV
for a sample of 1665 candidates.
The \Zmm candidates have one tight muon
and a second loose or tight muon.
Kinematic cuts of $\PT^{\mu1} \ge$ 20 and $\PT^{\mu2} \ge 15$ GeV
yielded 77 candidates.

Muon efficiencies are determined from the \Zmm sample.
The \Wmn and the \Zmm backgrounds are estimated from
data for the QCD and cosmic channels, and from Monte Carlo
for the \Zmm, \Ztt, \Wtm, and Drell-Yan backgrounds.
The luminosity for this trigger was $11.1 \pm 1.3 \ipb$
and the preliminary cross-sections are reported in table \ref{tab:xsecs}.
The electron and muon cross-sections are compared to
other measurements in figure 1.

\section{Ratio Measurements and \gw}

The ratio $R=\sbr(\Wln)/\sbr(\Zll)$
is of interest since it can be expressed as
the following combination of precisely measurable or calculable quantities:
\begin{equation}
 R \equiv \frac{\sigma B(W\to\mu\nu)}{\sigma B (Z\to\mu\mu)} =
\frac{\sgw}{\sgz} \hskip3mm \frac{\gmw}{\gmz} \hskip3mm \frac{\gz}{\gw}.
\nonumber
\label{eq:r_long}
\end{equation}

The measured value for the $Z$ width is obtained from the LEP
experiments~\cite{lepz}
\[ \gz = 2.487 \pm 0.010 \; \mbox{GeV/c$^2$}. \]
The ratio of the $W$ and $Z$ leptonic decay widths is taken from its
theoretical
value~\cite{hollik}
\[ \frac{\gmw}{\gmz} = 2.70 \pm 0.01. \]
The ratio of the $W$ to $Z$ production is determined using
the complete $O(\alpha^2_s)$ calculation~\cite{xsec_theo}
convoluted with various parton distribution functions~\cite{D0N1462}
to obtain
\[ \frac{\sgw}{\sgz} = 3.34 \pm 0.03 \]
where the quoted error is dominated by
the uncertainty on the $W$ mass and
systematic differences in the structure functions.

We take the weighted average of the muon and electron channels'
measurements of $R$:
\begin{eqnarray}
R^e     &=& 10.54 \pm 0.39 (stat) \pm 0.55 (syst), \nonumber \\
R^{\mu} &=& 12.0 ^{+1.8}_{-1.4} (stat) \pm 1.0 (syst), \nonumber \\
R^{e\mu}&=& 10.78 ^{+0.68}_{-0.60} (stat+syst). \nonumber
\end{eqnarray}
Combining this average value with
equation~\ref{eq:r_long} yields the total width of the $W$
\[ \gw = 2.08 \pm 0.12(stat + syst) \pm 0.02(theory+LEP syst) \;
\mbox{GeV}. \]

This result can be compared with the Standard Model prediction~\cite{smgw} of
\[ \gw = 2.09 \pm 0.02 \; \mbox{GeV} \]
for $M_t > M_b + M_W$ where $M_t$, $M_b$ and $M_W$ are the masses of the top
quark, bottom quark and $W$ boson, respectively.

We can set a limit on the decay of the $W$ into new quark pairs.
If the $W$ couples to a new quark and to the b quark
with standard model coupling,
a limit on the mass of this quark is set:
\[ m_{q} > 56 \; \mbox{GeV at 95\% CL}. \]
This limit applies independent of any assumptions
of decay modes of the top quark
and is illustrated in figure 2.

\begin{figure}[htb]
\vskip -1.3in
\centerline{
\hskip  0.0in
\psfig{figure=bandw_world_wzxsec.ps,height=9cm,width=9cm}
\hskip -0.5in
\psfig{figure=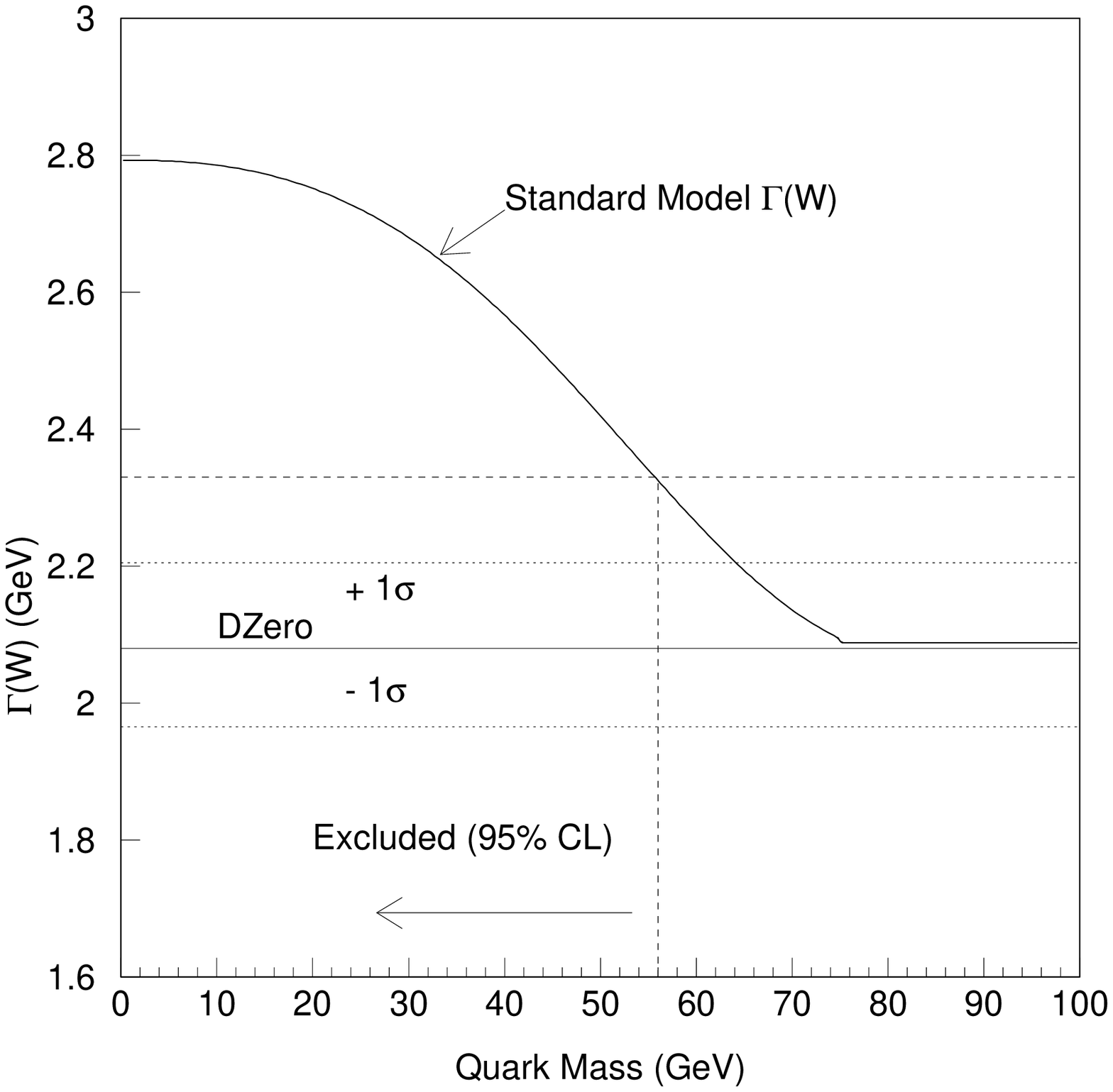,height=5.7cm,width=7.7cm}}
\label{fig:xsecs}
\caption{W and Z cross-sections \hskip 1.0in Fig. 2. Quark Mass Limit}
\end{figure}

\section{Summary}

\D0 has measured the production cross-sections for $W$ and $Z$
in \PAP\ at \rstev. We report preliminary results for those
cross-sections and for their ratio,
$R^{e\mu} = 10.78 ^{+0.68}_{-0.60}$. This yields
a measurement of the width of the $W$,
$\gw = 2.08 \pm 0.12 (exp) \pm 0.02 (thy)$ GeV,
and a limit on a new quark
$m_{q} > 56$ GeV at 95\% confidence level.

\end{document}